\documentclass{article}
\usepackage{spconf,amsmath,graphicx}
\usepackage{soul}
\usepackage{multirow}
\usepackage{url}
\usepackage{verbatim}

\title{Audiovisual speaker conversion: jointly and simultaneously \\transforming facial expression and acoustic characteristics}
\name{Fuming Fang$^1$, Xin Wang$^1$, Junichi Yamagishi$^{1,2}$, Isao Echizen$^1$}
\address{$^1$National Institute of Informatics, Tokyo, Japan \\
$^2$The University of Edinburgh, Edinburgh, UK}

\begin{document}
\ninept
\maketitle
\begin{abstract}
An audiovisual speaker conversion method is presented for simultaneously transforming the facial expressions and voice of a source speaker into those of a target speaker. Transforming the facial and acoustic features together makes it possible for the converted voice and facial expressions to be highly correlated and for the generated target speaker to appear and sound natural. It uses three neural networks: a conversion network that fuses and transforms the facial and acoustic features, a waveform generation network that produces the waveform from both the converted facial and acoustic features, and an image reconstruction network that outputs an RGB facial image also based on both the converted features.
The results of experiments using an emotional audiovisual database showed that the proposed method achieved significantly higher naturalness compared with one that separately transformed acoustic and facial features.
\end{abstract}
\section{Introduction}
\label{sec:intro}
With the development of information processing technology and the spread of Internet, voice and facial expression-based methods are being used more and more in our everyday lives. Two prominent methods are voice conversion~\cite{abe1990voice} and face transformation~\cite{thies2016face2face}, which change a person's voice or facial expressions into that or those of another person. These methods can be used for privacy protection, film/animation production, games, and other voice/facial signal-based transformations.

Changing both voice and facial expressions is important for certain applications, such as video games. An obvious way to achieve this is to separately transform the voice and facial expressions using two separate methods. This approach can result, however, in loss of naturalness due to asynchronous voice and facial movements due to transformation errors, delays (when considering context information), and other factors. Naturalness can be improved by using a synchronization method.

Another way to improve naturalness is to utilize the correlation between speech and facial movements~\cite{ananthakrishnan2012exploring} and jointly transform them so that they are always associated together. We have developed such a method, an audiovisual speaker conversion (AVSC) method that simultaneously transforms acoustic and facial characteristics. It uses three neural networks: a conversion network that fuses and transforms the acoustic and facial features of a source speaker into those of a target speaker, a waveform generation network that produces the waveform given both the converted acoustic and facial features, and an image generation network that outputs the RGB facial image also based on both the converted features. With the proposed method, we observed higher naturalness and quality than when the acoustic and facial features were separately transformed. This appears to be the first ever research on integrating voice conversion and face transformation in one system\footnote{A demonstration is available at \url{https://nii-yamagishilab.github.io/avsc/index.html}}.

The rest of this paper is organized as follows. Section~\ref{sec:relatedwork} discusses the differences between the proposed method and related ones. Section~\ref{sec:proposal} describes the proposed method. Section~\ref{sec:experimental_setup} describes the experimental conditions, and Section \ref{sec:results} presents and discusses the results. Finally, Section~\ref{sec:conclusion_futurework} summarizes the key points and mentions future work.

\vspace{-2mm}
\section{Related work}
\label{sec:relatedwork}
\vspace{-1mm}
The proposed method is related to work in four areas: audiovisual voice conversion, audiovisual speech enhancement, lip movement-to-speech synthesis, and speech-to-lip movement synthesis.

Tamura et al.~\cite{satoshi2018audio} proposed an audiovisual voice conversion method that learns highly correlated acoustic and lip movement features by using deep canonical correlation analysis~\cite{andrew2013deep} and then ties them together as a new feature. They reported significant improvement in terms of speech quality under noisy conditions compared with using acoustic feature only. Gabbay et al.~\cite{gabbay2017visual} proposed an audiovisual speech enhancement method that fuses lip feature and a noisy spectrum using a neural network and directly predicts a clean spectrum. Gogate et al.~\cite{gogate2018dnn} and Afouras et al.~\cite{afouras2018conversation} developed similar speech enhancement methods that predict a mask from lip images and noisy acoustic features using a neural network to filter out noise.

A lip movement-to-speech synthesis system was developed by Kumar et al.~\cite{Kumar2018Harnessing} that pairs mouth image sequence obtained using multi-view cameras with the corresponding audio to learn a mapping function. Kumar et al.~\cite{kumar2017obamanet} and Suwajanakorn et al.~\cite{suwajanakorn2017synthesizing} developed similar speech-to-lip movement synthesis methods that generate mouth keypoints from audio information and then render RGB images of the mouth from the mouth shape (represented by the keypoints). Taylor et al.~\cite{taylor2017deep} used speech to control an avatar.

The proposed method differs from these methods in that it uses not only lip movements but also facial expressions and movements. Furthermore, it transforms both audio and facial expressions.

\vspace{-2mm}
\section{Proposed method}
\label{sec:proposal}
\begin{figure}[tb]
\begin{center}
    \includegraphics[width=0.8\linewidth]{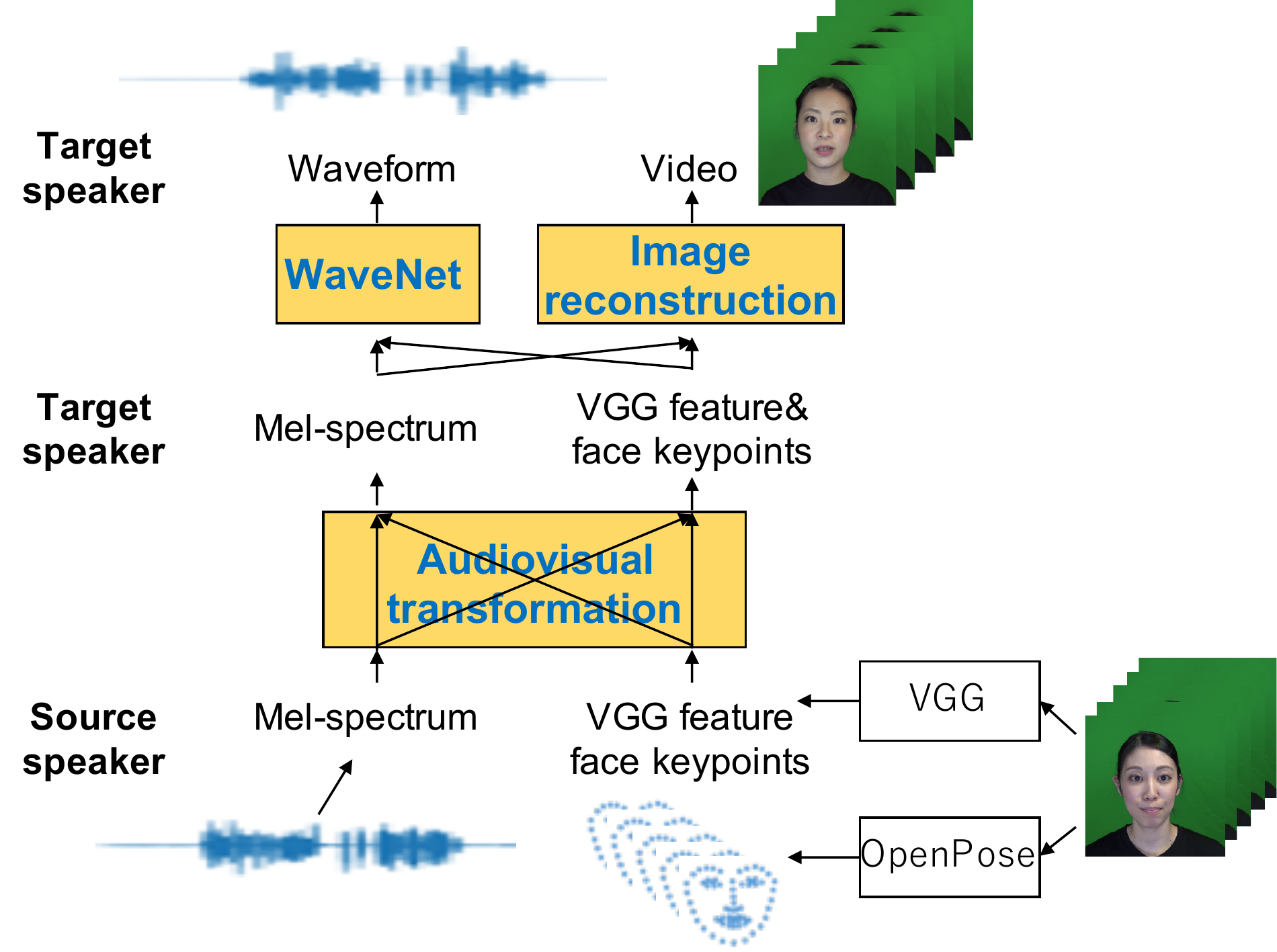}
\end{center}
\vspace{-6mm}
\caption{An implementation of proposed audiovisual speaker conversion method.}
\vspace{-4mm}
\label{fig:proposal}
\end{figure}

\vspace{-1mm}
The main idea of the proposed AVSC method is to correlate acoustic and facial characteristics so that they can compensate for each other during transformation and achieve high naturalness. Figure~\ref{fig:proposal} illustrates an implementation of the proposed method, which uses three networks: an audiovisual transformation network, a WaveNet~\cite{oord2016wavenet}, and an image reconstruction network. The audiovisual transformation network is used to convert acoustic and facial features from a source speaker to a target speaker. The WaveNet and the image reconstruction network synthesize speech and RGB images from the transformed acoustic and facial features, respectively. Finally, a video of the target speaker is created using the synthesized speech and image sequence.

\vspace{-2mm}
\subsection{Audiovisual transformation network}
\vspace{-1mm}
The acoustic feature is the mel-spectrum with 80 dimensions. It is extracted from the waveform by setting the window size to 25 ms and the hop length to 5 ms. The facial feature is extracted using a pre-trained VGG-19~\cite{simonyan2014very} network. The 17th hidden layer's output is used (denoted as ``VGG feature''). Because the VGG-19 is stacked by convolutional neural network (CNN)~\cite{lecun1998gradient} and max-pooling layers followed by fully connected layers, most of the original geometric information is lost, and only high-level features are preserved. To enable facial geometric information to be used, facial keypoints are extracted from each video frame by using OpenPose~\cite{cao2017realtime}. These keypoints describe the location and shape of a face. For example, they mark the contours of the jaw, mouth, nose, eyes, and eyebrows. The VGG feature (4096 dimensions) and face keypoints ($70 \times 2 = 140$ dimensions) are concatenated to create a new facial feature (4236 dimensions). In addition, since video data has a frame rate of 25 fps (40 ms per frame), a facial feature corresponds to eight mel-spectrum frames.

\begin{figure}[tb]
\begin{center}
    \includegraphics[width=0.95\linewidth]{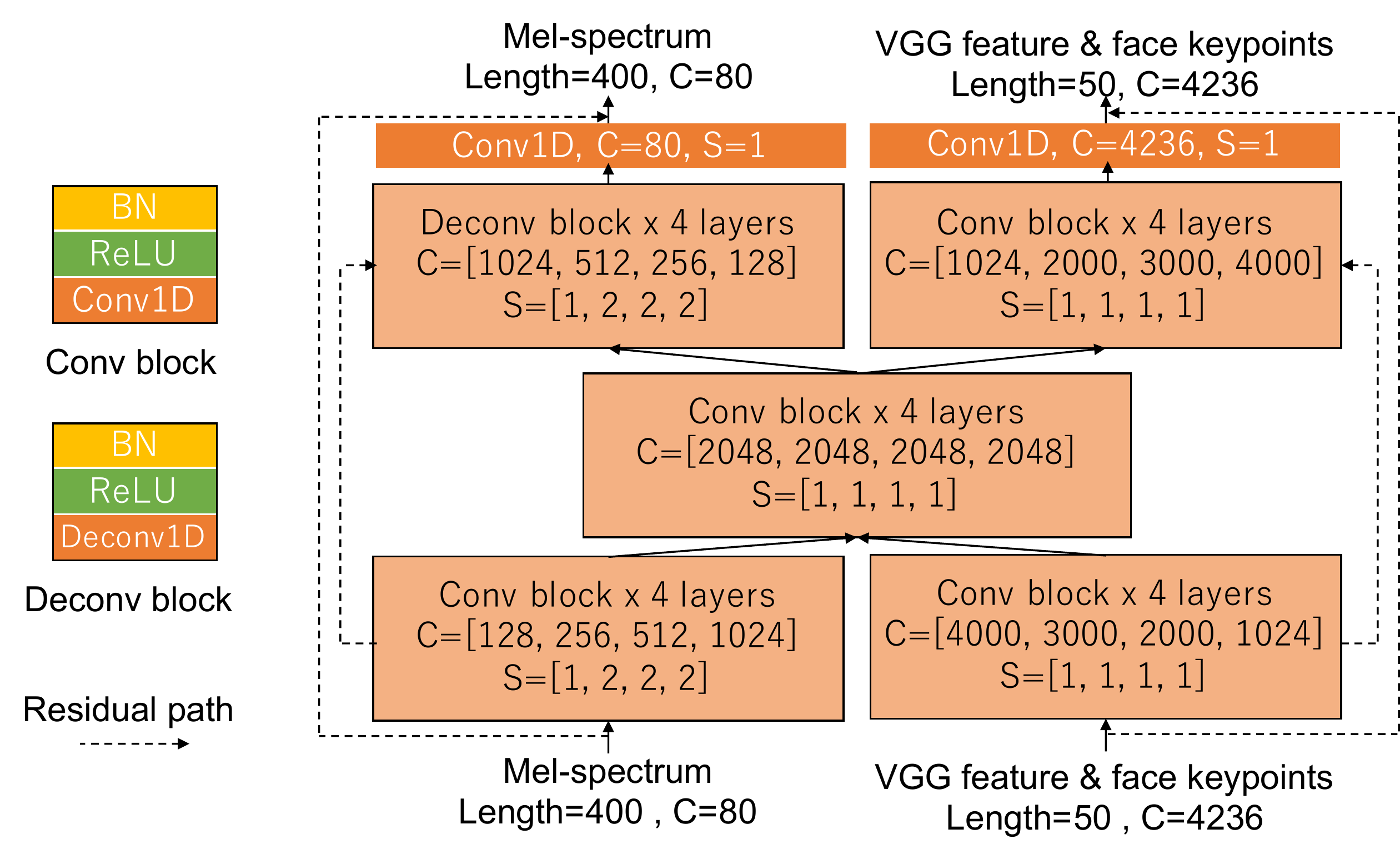}
\end{center}
\vspace{-6mm}
\caption{Audiovisual transformation network. ``BN'' denotes batch normalization, ``Conv1D'' denotes 1-D CNN layer, ``Deconv1D'' means transposed 1-D CNN layer, ``C'' means number of channels for each layer, and ``S'' means stride for each layer. Kernel size for all convolution layers is five. Activation function of hidden layers is ReLU and that of output layers is linear function. Batch normalization is performed in all hidden layers and not in output layer. Residual path connects feature maps having same shape.}
\vspace{-4mm}
\label{fig:audiovisual_network}
\end{figure}

Figure~\ref{fig:audiovisual_network} shows the architecture of the audiovisual transformation network, which contains a stack of 1-D convolutional layers. Its design was inspired by the work of Afouras et al.~\cite{afouras2018conversation}. The convolutions are performed along the temporal dimension, and the feature dimension is treated as channel. This makes it possible to match the sampling rates for the acoustic and facial features by adjusting the stride. This design also enables context information to be taken into account and thereby generate more fluent audio and facial movements. The network consists of five sub-networks and two output layers. Based on sampling rate of the facial feature, The lower left sub-network down-samples the acoustic feature by performing convolution three time with a stride of two. The lower right one maintains the facial feature sampling rate and reduces the feature dimension. The middle one fuses acoustic and facial information and associates them together. The upper left one up-samples and transforms the fused features into the target speaker's acoustic feature domain using transposed convolution~\cite{zeiler2010deconvolutional} layers. The upper right one transforms the fused features into target speaker's facial feature domain. In addition, the feature maps having the same shape are connected by a residual path~\cite{he2016deep}. Batch normalization~\cite{Ioffe2015batch} is performed in each hidden layer after rectified linear unit (ReLU)~\cite{xu2015empirical} activation.

The training data is truncated every two seconds, so the mel-spectrum and facial feature have shapes of $400\times 80$ and $50\times 4236$, respectively. The L1 norm is used as the training objective, and the acoustic part loss is weighted by 10. During the test phase, the entire length of test data was input to the network.

\vspace{-2mm}
\subsection{WaveNet}
\vspace{-1mm}
The WaveNet is used to convert the transformed mel-spectrum and facial feature into the speech waveform. WaveNet is an autoregressive~\cite{jordan1997serial} neural-network-based waveform model that generates waveform sampling points one by one. 

The WaveNet structure is the same as that of the one used in another study~\cite{luong2018investigating} except the condition module which takes in the mel-spectrum as input. It consists of a linear projection layer, 40 dilated convolution~\cite{YuKoltun2016} layers, a post-processing block with a softmax output layer, and a condition module. The linear projection layer takes as input a waveform value generated in the previous time step while the condition module takes the transformed mel-spectrum and facial feature as input. Given the outputs from the linear layer and the condition module, the dilated convolution layers compute hidden features, which the post-processing module uses to compute the distribution waveform sampling point for the current time step. A waveform value is generated from this distribution, and this process is repeated to generate the entire waveform.

\vspace{-2mm}
\subsection{Image reconstruction network}
\begin{figure}[tb]
\begin{center}
    \includegraphics[width=0.7\linewidth]{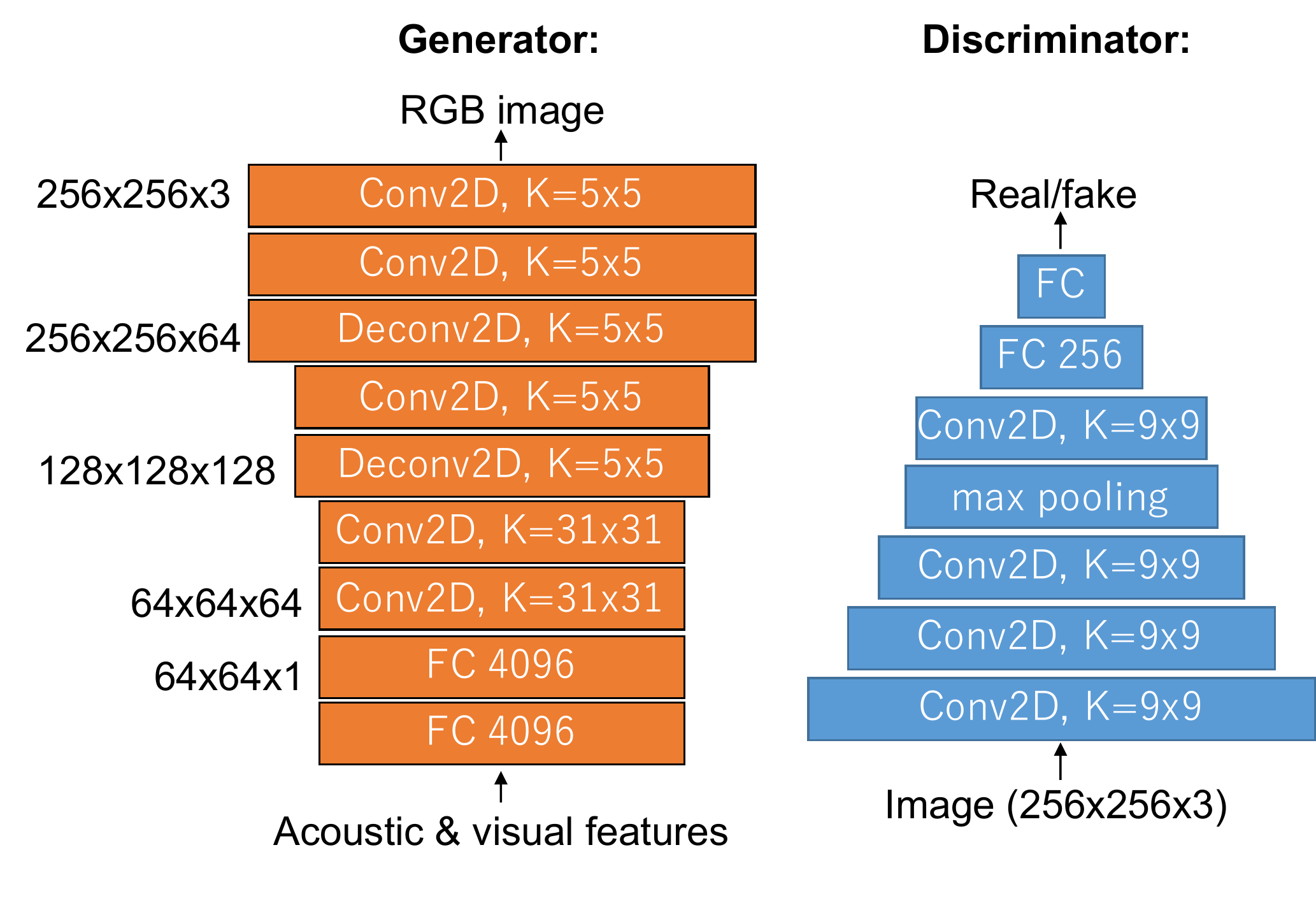}
\end{center}
\vspace{-10mm}
\caption{Image reconstruction network. Generator is used to reconstruct image from acoustic and visual features. Discriminator is an additional network used for training generator using adversarial loss. ``Conv2D'' means 2-D CNN layer, ``Deconv2D'' means transposed 2-D CNN layer, ``FC'' means fully connected layer, and ``K'' means kernel size. The convolution layers in the generator have (from the bottom) 64, 64, 128, 128, 64, 64, and 3 channels. The Conv2D layers have a stride of one, and the Deconv2D layers have a stride of two. The Conv2D layers in the discriminator have (from the bottom) 8, 16, 32, and 32 channels. Both the Conv2D and max pooling layers have a stride of two. Activation function of all hidden layers is ReLU. Batch normalization is performed in the generator before activation and not in the discriminator.}
\label{fig:imageconstruction}
\vspace{-4mm}
\end{figure}

\vspace{-1mm}
The image reconstruction network synthesizes an RGB image from the transformed acoustic and facial features. It contains nine stacked layers: two fully connected layers and seven convolution layers, as shown on the left in Figure~\ref{fig:imageconstruction}. Eight frames of acoustic features and one frame of facial feature are concatenated and input. The network first transforms the concatenated feature into a fused feature with 4096 dimensions using two fully connected layers. The fused feature is then reshaped into a 2-D image ($64\times64\times1$) and sent to the next convolution layer. Finally, an image ($256\times256\times3$) is generated by performing convolutions and transposed convolutions.

Training of the image reconstruction network is based on a least squares generative adversarial network (LSGAN)~\cite{mao2017least} consisting of a generator and a discriminator (Figure~\ref{fig:imageconstruction} right side). The discriminator maximizes the probability of data from training images and minimizes the probability of generated images from the generator. The generator strives to generate images similar to the training data in order to maximize the probability and fool the discriminator. The L1 norm (weighted by 10) is used to stabilize the training process.

\section{Experimental setup}
\label{sec:experimental_setup}
\vspace{-1mm}
We compared the performance of the proposed AVSC method with a baseline method that separately transforms acoustic and facial features. We carried out an objective and a subjective experiment. The objective experiment evaluated the correlation between speech and lip movements. The subjective experiment evaluated the naturalness, quality, and speaker similarity of the converted speech and video.

\vspace{-2mm}
\subsection{Database}
\vspace{-1mm}
To accurately evaluate the correlation between audio and facial features, we created an emotional audiovisual database using input from two Japanese female actors. Seven emotions were defined: neutral, normal happiness, strong happiness, normal sadness, strong sadness, normal anger, and strong anger. For each emotion, we used 100 different sentences selected from dialogs in novels. We asked the two actors to utter each sentence while displaying the corresponding emotion. Four people monitored the recording sessions, and if any of them felt that the target emotion was not displayed in the speech or facial expression, the actor was instructed to repeat the recording of that sentence.

The recording took place in a soundproof chamber. A condenser microphone (NEUMANN87) and a video camera (Sony HDR-720V/B) were set at front of the actor. A green cloth sheet was used as the background. The audio was recorded at 96 kHz with 24-bit resolution. The video was recorded at 60 fps with $1920\times1080$ resolution. There were 17 recording sessions in total. A clapperboard was clapped shut at the beginning of each session to enable the audio to be synchronized with the video. Finally, the audio signal recorded by the video camera was replaced with that recorded using the condenser microphone.

Since the two actors used the same sets of sentences, the database had parallel recordings (700 per actor). The duration was approximately 1h10min for each actor. The average sentence duration was 5.9 s. We refer to the two actors as speakers F01 and F02.

\vspace{-3mm}
\subsection{Training data and test data}
\vspace{-2mm}
We designated speaker F01 as the source speaker and F02 as the target speaker. We randomly selected 90 data samples for each emotion as training data (630 samples in total) for each speaker. The remaining 70 samples were used as test data. We down-sampled the audio signal to 48 kHz with 16-bit resolution and then extracted the mel-spectrum. We down-sampled the video signal to 25 fps, centered the speaker, and resized the video to $1080\times1080$. We then extracted the image and resized it to $224\times224$ for VGG feature extraction and $256\times256$ for face keypoint extraction as well as for use as training data for the image reconstruction network.

\vspace{-3mm}
\subsection{Proposed method setup}
\vspace{-2mm}
For dynamic time warping (DTW)-based alignment, the distance between features from the two speakers was calculated by summing the dimension-averaged L2-norm of the acoustic feature and that of the facial feature (we tied every eight acoustic features to match length of the facial feature). The learning rate for the audiovisual transformation network was $10^{-4}$, the mini-batch size was 64, and the number of epochs was 600.

The WaveNet was adapted from a pre-trained model~\cite{luong2018investigating} that was trained using 15 hours of Japanese speech data. We fine tuned over 199 epochs using the F02 training data. The number of epochs was set on the basis of the results of a preliminary experiment. 

The image reconstruction network was trained using the F02 training data. The learning rates were set to $10^{-3}$ and $10^{-5}$ for the generator and discriminator, respectively, the mini-batch size was 64, and number of epochs was 30.

\vspace{-3mm}
\subsection{Baseline setup}
\vspace{-2mm}
For the baseline method, we removed the acoustic-related or facial-related part from the proposed method. In addition, since it was not necessary for the baseline method to down- or up-sample acoustic features, we changed the stride of the audiovisual transformation network to one. From the results of a preliminary experiment on WaveNet adaptation, we set the number of epochs to 100. The other hyper-parameters (learning rate, mini-batch size, and number of epochs) were the same as for the proposed method. As additional information, we tuned the learning rate and number of epochs using the baseline method and directly applied them to the proposed method. We did not tune mini-batch size.

\vspace{-3mm}
\subsection{Evaluation setup}
\vspace{-2mm}
We evaluated the correlation between the converted speech and lip movements using canonical correlation analysis, which calculates correlation coefficient $r$ between two sequences. We re-extracted the mel-spectrum and lip keypoints from the converted speech and images. We set the window length and hop length for the mel-spectrum to 40 ms and 40 ms, respectively.

The naturalness and quality were evaluated on a 1-to-5 Likert mean opinion score (MOS) scale. The speaker similarity was evaluated using a preference test. The evaluation was carried out by means of a crowdsourced web-based interface. On each web page, we presented three questions about naturalness and quality for the audio-only evaluation case, the visual-only evaluation case, and the audiovisual evaluation case. We presented only audio or silent video for the audio-only evaluation case and the visual-only evaluation case. For the audiovisual evaluation case, we asked the evaluators to view a video and assess the quality of speech, image, and synchronization between speech and lip movement. We also presented additional three questions about speaker similarity for the audio-only evaluation case, visual-only evaluation case, and audiovisual evaluation case. The evaluators were limited to a maximum of 50 pages, and they had to listen/view all samples and answer all questions. There was a total of 186 valid evaluators, and they produced 4995 page data points, which is equivalent to 35.7 evaluations per sample.

The statistical significance analysis was based on an unpaired two-tail $t$-test with a 99\% confidence interval.

\vspace{-2mm}
\section{Results}
\label{sec:results}
\subsection{Correlation between speech and lip movements}
\begin{figure}[tb]
\begin{center}
    \includegraphics[width=0.9\linewidth]{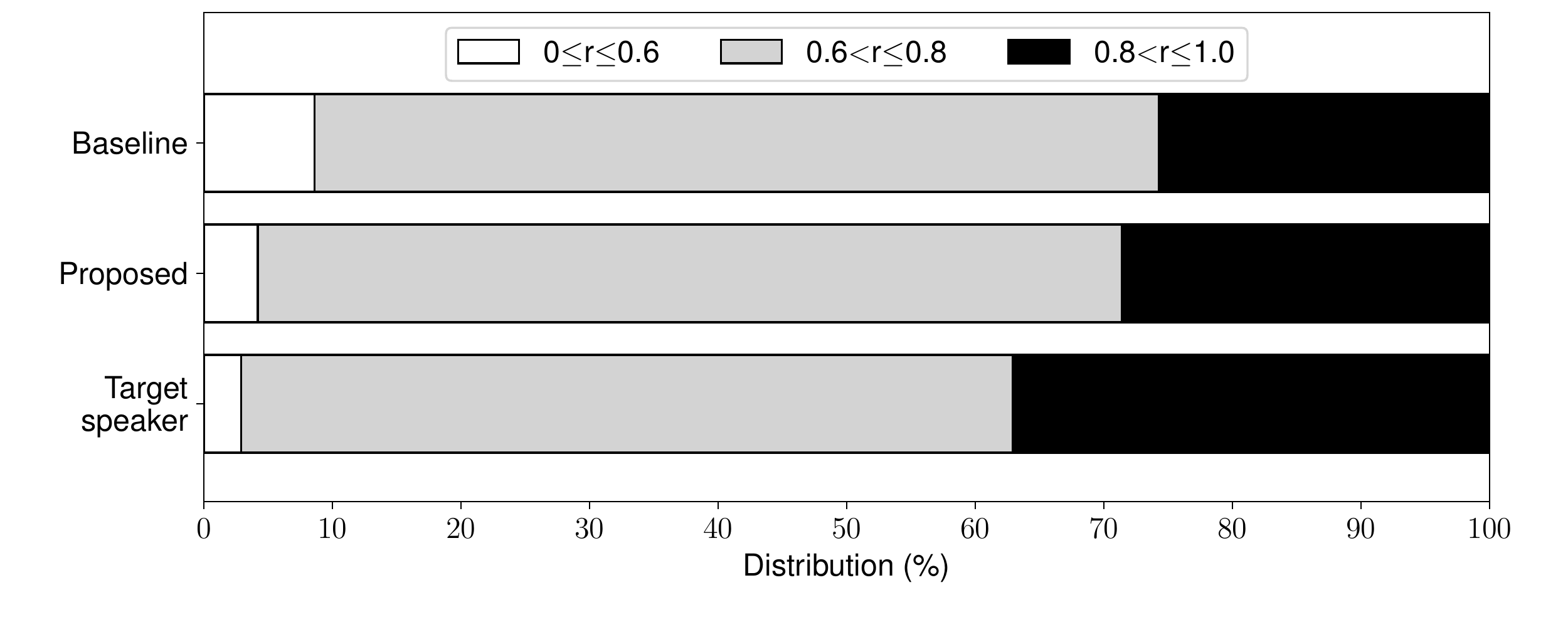}
\end{center}
\vspace{-10mm}
\caption{Distribution of correlation coefficient $r$ between mel-spectrum and lip movements.}
\label{fig:cca}
\vspace{-4mm}
\end{figure}

\vspace{-2mm}
Although our main intention is to correlate voice and facial expression, the proposed method includes synchronized generation of speech and lip movements.
Figure~\ref{fig:cca} shows the distribution of correlation coefficients between the mel-spectrum and lip movements. The correlation was higher for the proposed method than for the baseline. This suggests that acoustic and facial feature are more closely associated if they are jointly and simultaneously transformed. However, there was a gap between the results for the proposed method and for the target speaker.

\vspace{-3mm}
\subsection{Subjective evaluation}
\vspace{-2mm}
As shown in Table~\ref{tbl:quality}, the MOS values with the proposed method were significantly better in both the audio-only and audiovisual evaluation cases achieved than with the baseline method.
One reason was that the facial feature compensated for the acoustic feature and the proposed method achieved better synchronous.
The slightly better performance of the proposed method in the visual-only evaluation case is attributed to the facial feature dominating the fused feature, making it difficult to take advantage of the acoustic feature. The scores for the emotional test samples were smaller than those for the neutral samples in almost case with the baseline method. It was possible to achieve a higher or similar score as the neutral one by fusing both the acoustic and facial features. e.g., sadness in the audio-only evaluation case and happiness in the audiovisual evaluation case. This indicates that facial movements and some speech characteristics might help enhance emotion transformation.
\begin{table}[tb]
  \centering
  \caption{MOS values for speech, visual, and audiovisual naturalness/quality. ``Pr'' means proposed method, ``Bs'' means baseline method, and ``+'' indicates strong emotion.}
  \vspace{3pt}
  \label{tbl:quality}
  \begin{tabular}{|c|cccccc|}
  \hline
    \multirow{3}{*}{Emotion type}& \multicolumn{6}{c|}{Evaluation modality} \\ \cline{2-7}
    & \multicolumn{2}{c}{Audio-only} & \multicolumn{2}{c}{Visual-only} & \multicolumn{2}{c|}{Audiovisual} \\ 
    & Pr & Bs & Pr & Bs & Pr & Bs \\
    \hline
    Neutral & 2.70 & 2.33 & 3.79 & 3.65 & 3.37 & 2.98 \\
    Happiness & 2.39 & 2.15 & 3.69 & 3.73 & 3.41 & 2.93 \\
    +Happiness & 2.33 & 2.05 & 3.36 & 3.33 & 3.25 & 2.85 \\
    Sadness & 2.72 & 2.24 & 3.32 & 3.40 & 3.21 & 2.95 \\
    +Sadness & 2.24 & 2.01 & 3.46 & 3.45 & 3.31 & 3.16 \\
    Anger & 2.46 & 2.08 & 3.42 & 3.35 & 3.22 & 2.99 \\
    +Anger & 1.96 & 1.79 & 3.27 & 3.11 & 3.06 & 2.76 \\
    \hline
    Average & 2.40 & 2.09 & 3.47 & 3.43 & 3.26 & 2.95 \\
    \hline
  \end{tabular}
\end{table}
\begin{figure}[t]
\begin{center}
    \includegraphics[width=1.0\linewidth]{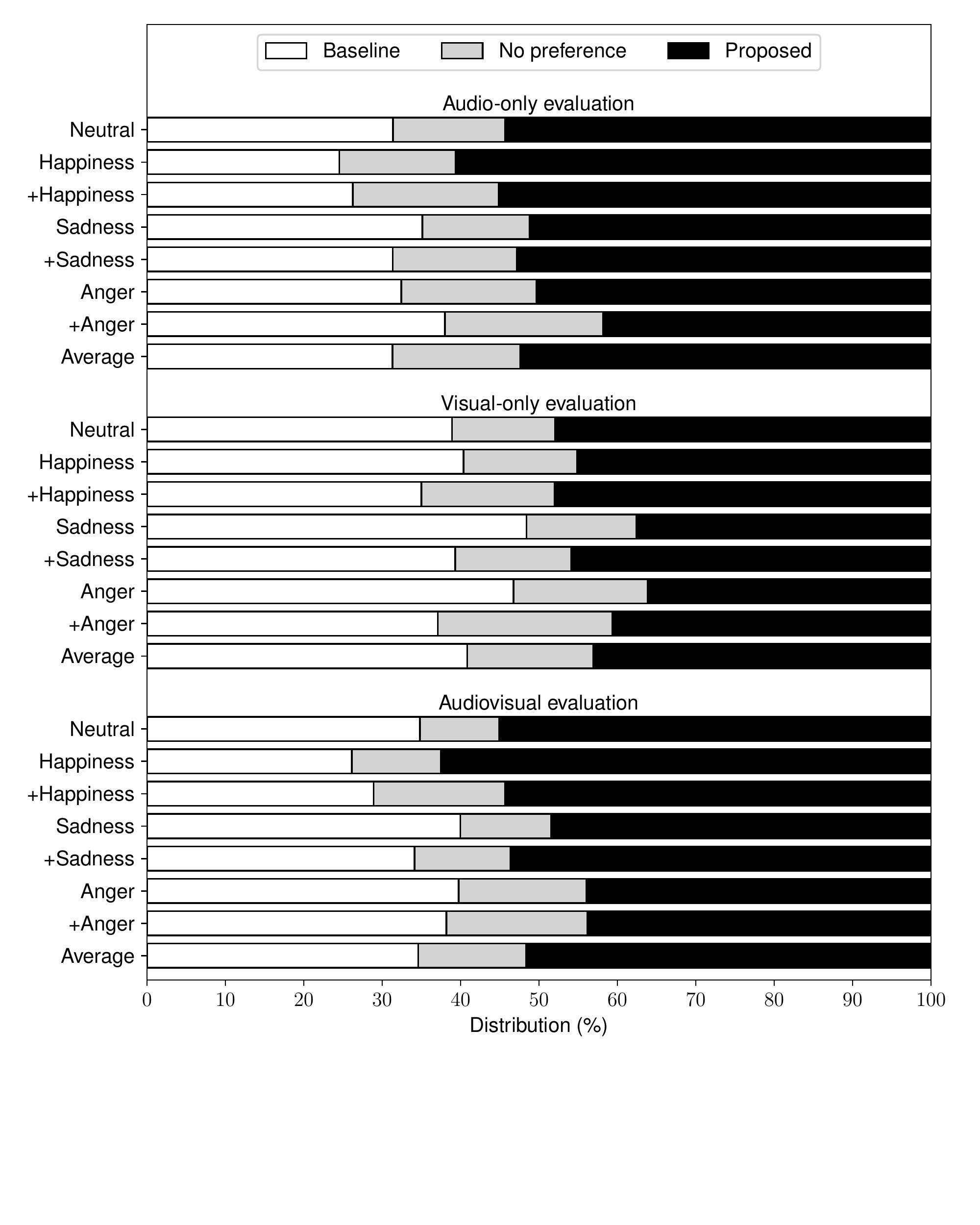}
\end{center}
\vspace{-22mm}
\caption{Speaker similarity preference test results. ``+'' means strong emotion.}
\label{fig:preference}
\vspace{-6mm}
\end{figure}

The preference test results (Figure~\ref{fig:preference}) were similar. The proposed method achieved higher speaker similarity than the baseline method for the audio-only and audiovisual evaluation cases. This might be because facial identity and voice identity helped to estimate accurate parameters of the networks.

\vspace{-3mm}
\section{Summary and future work}
\label{sec:conclusion_futurework}
\vspace{-2mm}
Our proposed audiovisual speaker conversion method simultaneously transforms voice and facial expression of a source speaker into those of a target speaker. We implemented this method using three networks: a conversion network fuses and transforms the acoustic and facial features of the source speaker into those of the target speaker, a WaveNet synthesizes the waveform from the converted features, and an image reconstruction network generates RGB images from the converted features. Experiments using an emotional audiovisual database showed that the proposed method can achieve higher naturalness/quality and speaker similarity than a baseline method that separately transforms the acoustic and facial features.

Since the facial features may dominate the transformation, we plan to improve our method to better balance the acoustic and facial features. The use of a parallel training approach makes it necessary to align training data, so we had to carefully balance the acoustic and facial features. We will thus consider developing a non-parallel training method~\cite{vc2018cyclegan} for audiovisual speaker conversion.

\vspace{-2mm}
\section{Acknowledgement}
\vspace{-1mm}
This work was supported by JSPS KAKENHI Grant Numbers (16H06302, 17H04687, 18H04120, 18H04112, 18KT0051) and by JST CREST Grant Number JPMJCR18A6, Japan.

\bibliographystyle{IEEEbib}
\bibliography{refs}

\newpage
\section{Appendix}
\begin{figure*}[h]
\begin{center}
    \includegraphics[width=0.9\linewidth]{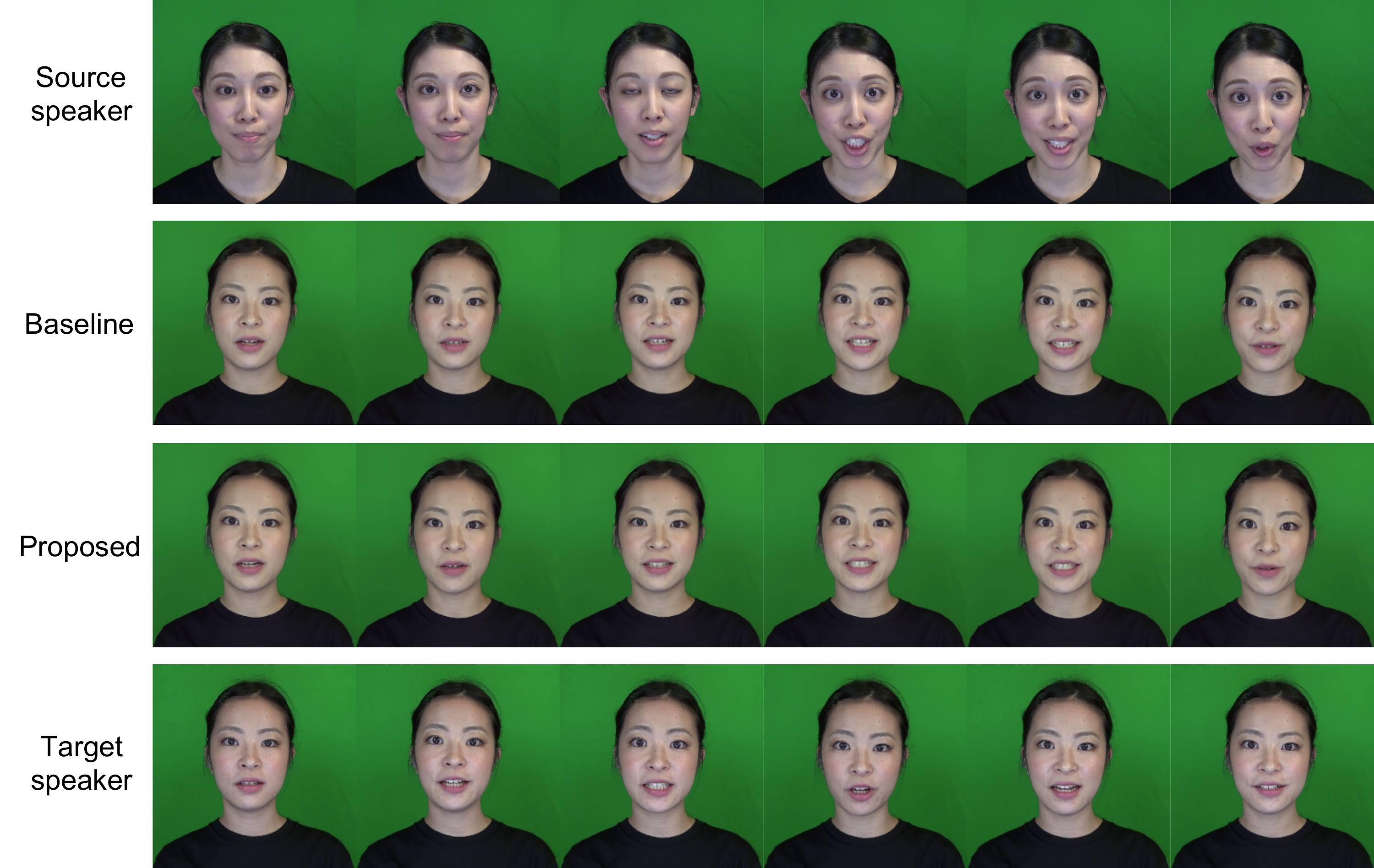}
\end{center}
\vspace{-5mm}
\caption{Example of converted images with neutral emotion.}
\label{fig:neutral}
\vspace{-4mm}
\end{figure*}

\begin{figure*}[h]
\begin{center}
    \includegraphics[width=0.9\linewidth]{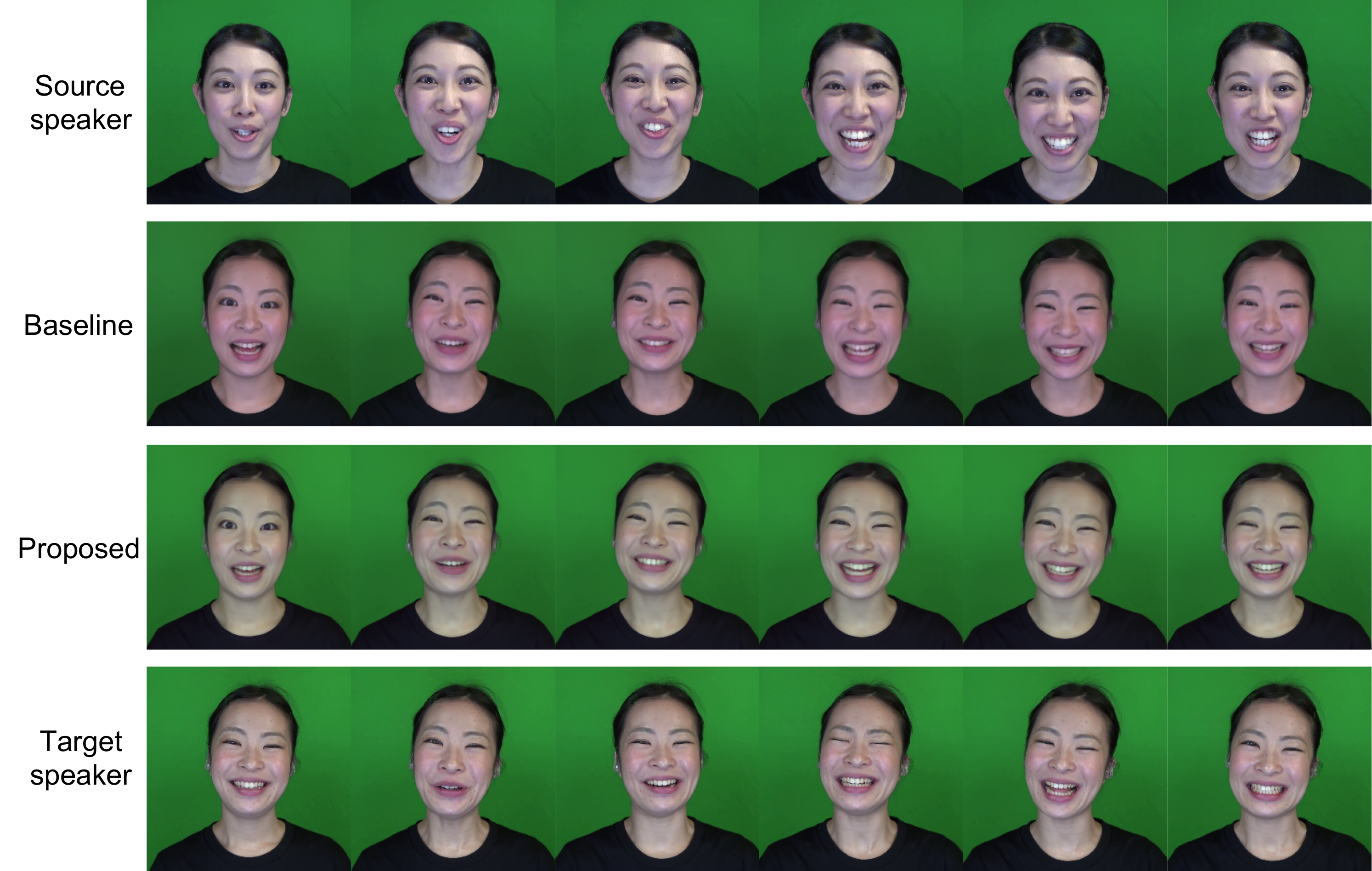}
\end{center}
\vspace{-5mm}
\caption{Example of converted images with strong happiness emotion.}
\label{fig:strong_happiness}
\vspace{-4mm}
\end{figure*}

Examples of images converted using the proposed and baseline methods are shown in Figure~\ref{fig:neutral}-\ref{fig:strong_happiness}. It is difficult to tell which method generated higher quality images or which one generated images similar to the target speaker. However, when the audio and video were played together (the audiovisual evaluation case), the proposed method had better synchronization between the audio and facial movements than the baseline. This means that the proposed method achieves better association between acoustic and facial features.

\begin{figure*}[h]
\begin{center}
    \includegraphics[width=0.9\linewidth]{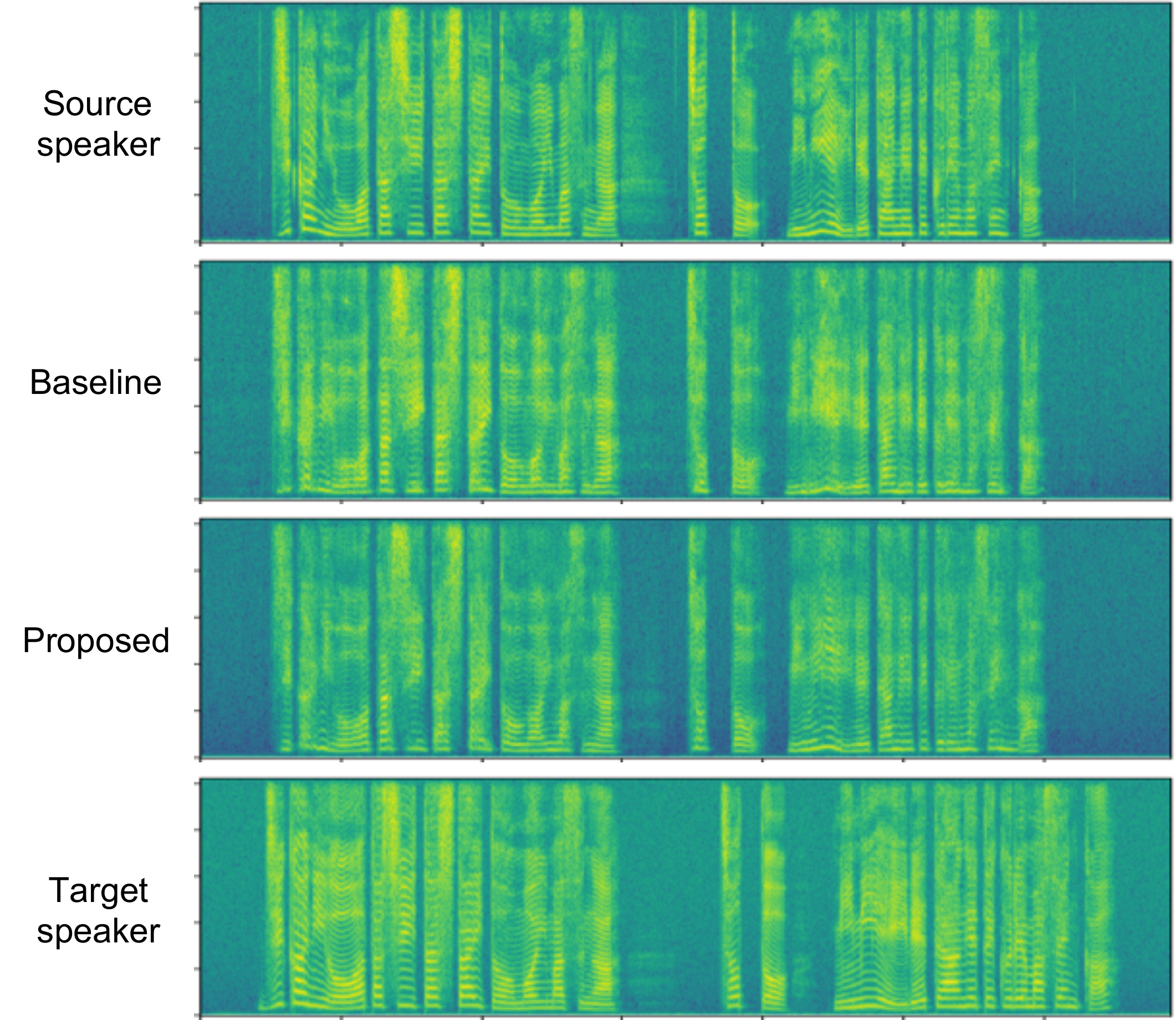}
\end{center}
\vspace{-5mm}
\caption{Example of converted spectrograms with neutral emotion.}
\label{fig:sp_neutral}
\vspace{-4mm}
\end{figure*}

\begin{figure*}[h]
\begin{center}
    \includegraphics[width=0.9\linewidth]{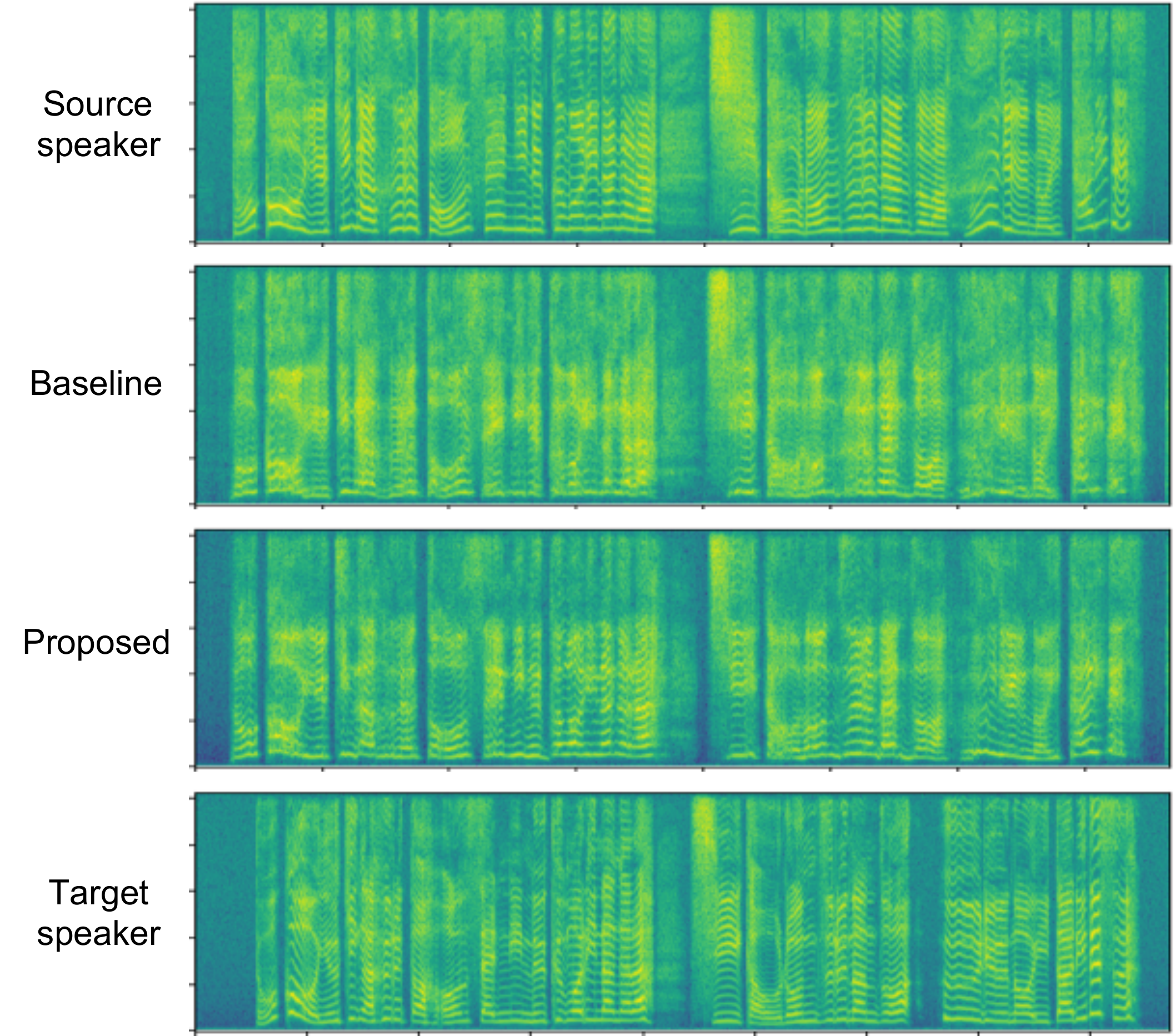}
\end{center}
\vspace{-5mm}
\caption{Example of converted spectrograms with strong happiness emotion.}
\label{fig:sp_strong_happiness}
\vspace{-4mm}
\end{figure*}

\begin{figure*}[h]
\begin{center}
    \includegraphics[width=0.9\linewidth]{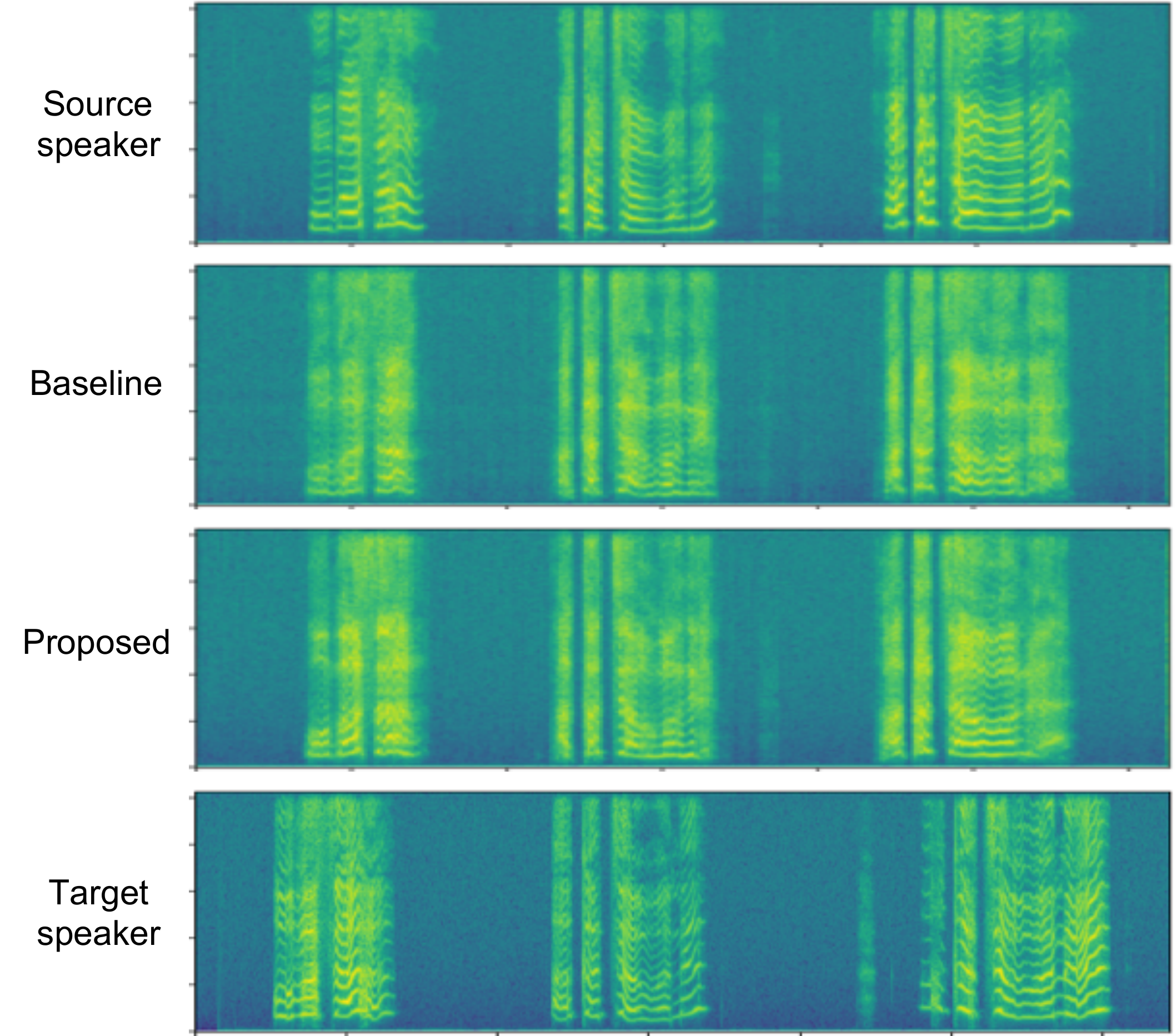}
\end{center}
\vspace{-5mm}
\caption{Example of converted spectrograms with strong sadness emotion.}
\label{fig:sp_strong_sadness}
\vspace{-4mm}
\end{figure*}

\begin{figure*}[h]
\begin{center}
    \includegraphics[width=0.9\linewidth]{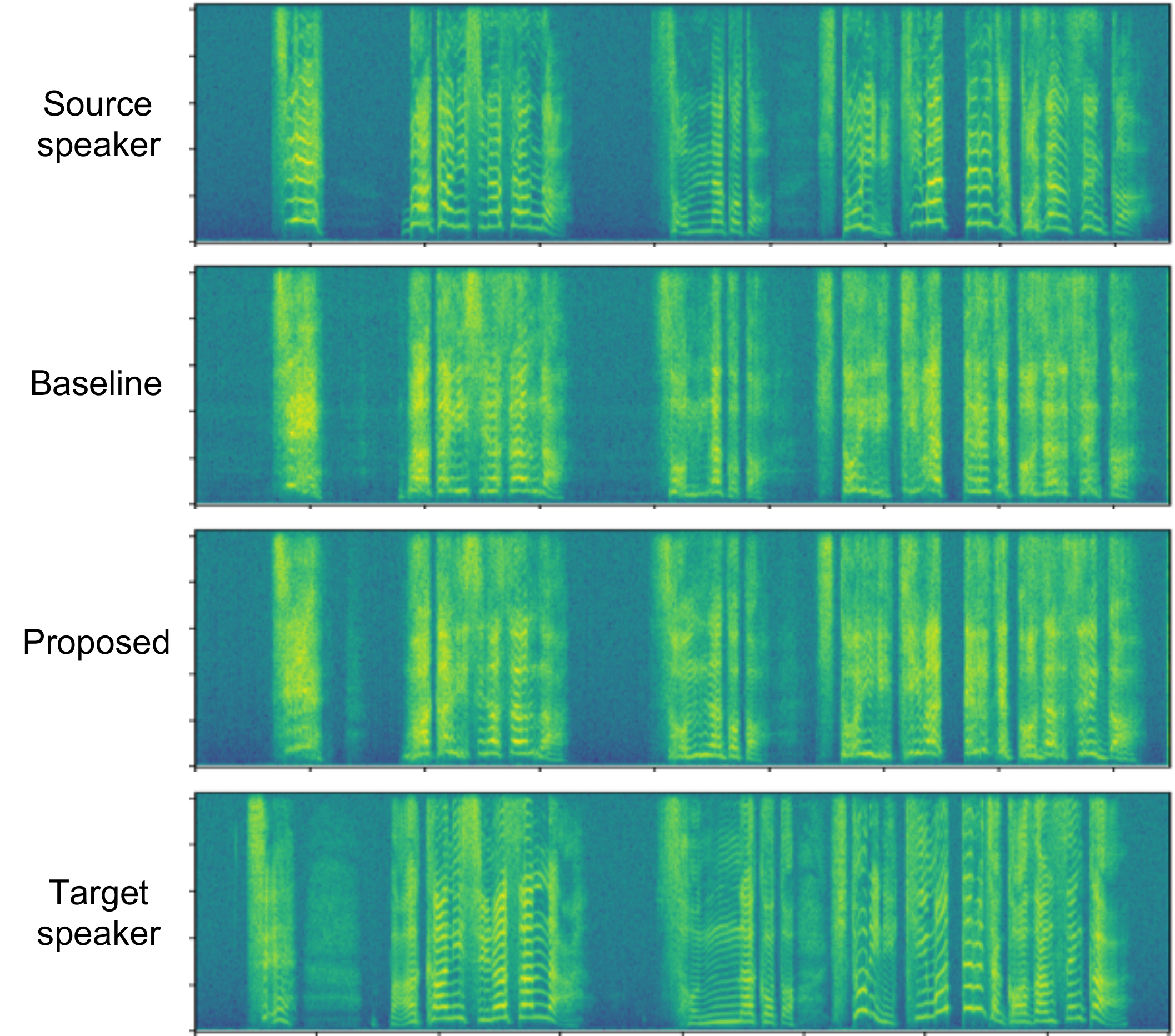}
\end{center}
\vspace{-5mm}
\caption{Example of converted spectrograms with strong anger emotion.}
\label{fig:sp_strong_anger}
\vspace{-4mm}
\end{figure*}

Examples of spectrograms converted using the proposed and baseline methods are shown in Figure~\ref{fig:sp_neutral}-\ref{fig:sp_strong_anger}. 
The horizontal direction indicates temporal axis while the vertical direction is frequency axis with range of 0 to 8000Hz.
It seems that the proposed method predicted better spectrogram than the baseline in most cases.

\end{document}